\documentclass[10pt,journal,doublecolumn,twoside]{IEEEtran}
\IEEEoverridecommandlockouts

\usepackage{subfigure}
\usepackage{colortbl}
\usepackage{bm}

\usepackage{graphicx}  
\usepackage{url}       
\usepackage{rotating}
\usepackage{amsmath}   
\usepackage{cite}

\usepackage{amsfonts,amssymb}

\usepackage{stfloats}

\usepackage{cases}
\usepackage{algorithm}
\usepackage{multirow}
\usepackage{algorithmic}

\newcommand{\myvec}[1]%
{\stackrel{\raisebox{-2pt}[0pt][0pt]
{\small$\rightharpoonup$}}{#1}}

\newcommand{\ls}[1]
    {\dimen0=\fontdimen6\the\font
     \lineskip=#1\dimen0
     \advance\lineskip.5\fontdimen5\the\font
     \advance\lineskip-\dimen0
     \lineskiplimit=.9\lineskip
     \baselineskip=\lineskip
     \advance\baselineskip\dimen0
     \normallineskip\lineskip
     \normallineskiplimit\lineskiplimit
     \normalbaselineskip\baselineskip
     \ignorespaces
    }

\begin{document}

\title{A Simple Channel Independent Beamforming Scheme with Circular Array}


\author{Haiyue Jing,~\IEEEmembership{Student Member,~IEEE}, Wenchi Cheng,~\IEEEmembership{Member,~IEEE}, and Xiang-Gen Xia,~\IEEEmembership{Fellow,~IEEE}



\thanks{\ls{.5}
Haiyue Jing and Wenchi Cheng are with the State Key Laboratory of Integrated Services Networks, Xidian University, Xi'an, 710071, China (e-mails: hyjing@stu.xidian.edu.cn and wccheng@xidian.edu.cn).

Xiang-Gen Xia is with Xidian University and the Department of Electrical and Computer Engineering, University of Delaware, Newark, DE 19716 (e-mail: xxia@ee.udel.edu).}
}


%


\maketitle
\thispagestyle{empty}
\pagestyle{empty}

\begin{abstract}
In this letter, we consider a uniform circular array (UCA) based line-of-sight (LOS) multiple-input-multiple-output (MIMO) system, where the transmit and receive UCAs are aligned with each other. We propose a simple channel independent beamforming scheme with fast symbol-wise maximum likelihood (ML) detection.
\end{abstract}


\section{Introduction}
\IEEEPARstart{D}{uring} the past two decades, multiple-input-multiple-output (MIMO) with uniform linear array (ULA) antenna architecture for wireless communications has been widely studied.
A number of efficient de/modulation schemes have been proposed in the literature, such as, Bell Laboratories Layered Space-Time (BLAST)~\cite{space_time_1996,BLAST_2003} and space-time block coded (STBC) schemes~\cite{space_time_1998,OSTBC_1999}. The demodulation methods include linear receivers, successful interference cancelation (SIC) receivers, and maximum-likelihood (ML) receivers. Although linear and SIC receivers are usually much faster than the ML receivers, their complexities are either at least in the third order of the signal block length or/and they have poor performances. To have a fast ML receiver, orthogonal STBC (OSTBC) etc. usually need to be equipped across several time slots. However, OSTBC may lead to a multiplexing loss~\cite{OSTBC_2003}.


Recently, uniform circular arrays (UCAs) have been recognized as a useful alternative for wireless communications over line-of-sight (LOS) channels~\cite{mmwave_UCA_2014}.
%
Because the antennas are uniformly placed around the perimeter of the circle, the channel matrix is a circulant matrix when the transmit and the receive UCAs are aligned with each other. Based on this property, in this letter we propose a simple channel independent beamforming scheme for the UCA based LOS MIMO system with the fast symbol-wise ML detection.
The idea is similar to the orthogonal frequency-division multiplexing (OFDM) system and instead of transmitting $N$ subcarrier signals across $N$ time slots in an OFDM system, it transmits them across $N$ transmit antennas in one time slot. It requires that $N$ receive antennas are equipped as well. Due to the fact that the channel matrix is circulant, the discrete Fourier transform (DFT)/inverse DFT (IDFT) can diagonalize the channel matrix similar to that for the intersymbol-interference (ISI) channel matrix after the cyclic prefix (CP) inserted/deleted in an OFDM system. After the diagonalization, the fast symbol-wise ML detection holds. We want to emphasize that our proposed scheme has symbol rate $N$, i.e., $N$ symbols per channel use.
We then derive the bit error rate (BER) for the LOS MIMO system. Also, we compare the BER and the number of computations for the LOS MIMO system using the channel independent beamforming with those of the traditional LOS MIMO system. Simulation results verify that our proposed channel independent beamforming scheme with the fast symbol-wise ML detection has the same BER performance as the traditional scheme, while the computational complexity is much lower.

This letter is organized as follows. In Section II, we describe the UCA model. In Section III, we present our simple channel independent beamforming scheme and its fast symbol-wise ML detection. We also present its BER and computational complexity analysis. In Section IV, we present some simulation results. In Section V, we conclude this letter.

\section{System Model for UCA Based MIMO Communications}\label{sec:sys}

\begin{figure}[t]
\centering
\includegraphics[width=0.45\textwidth]{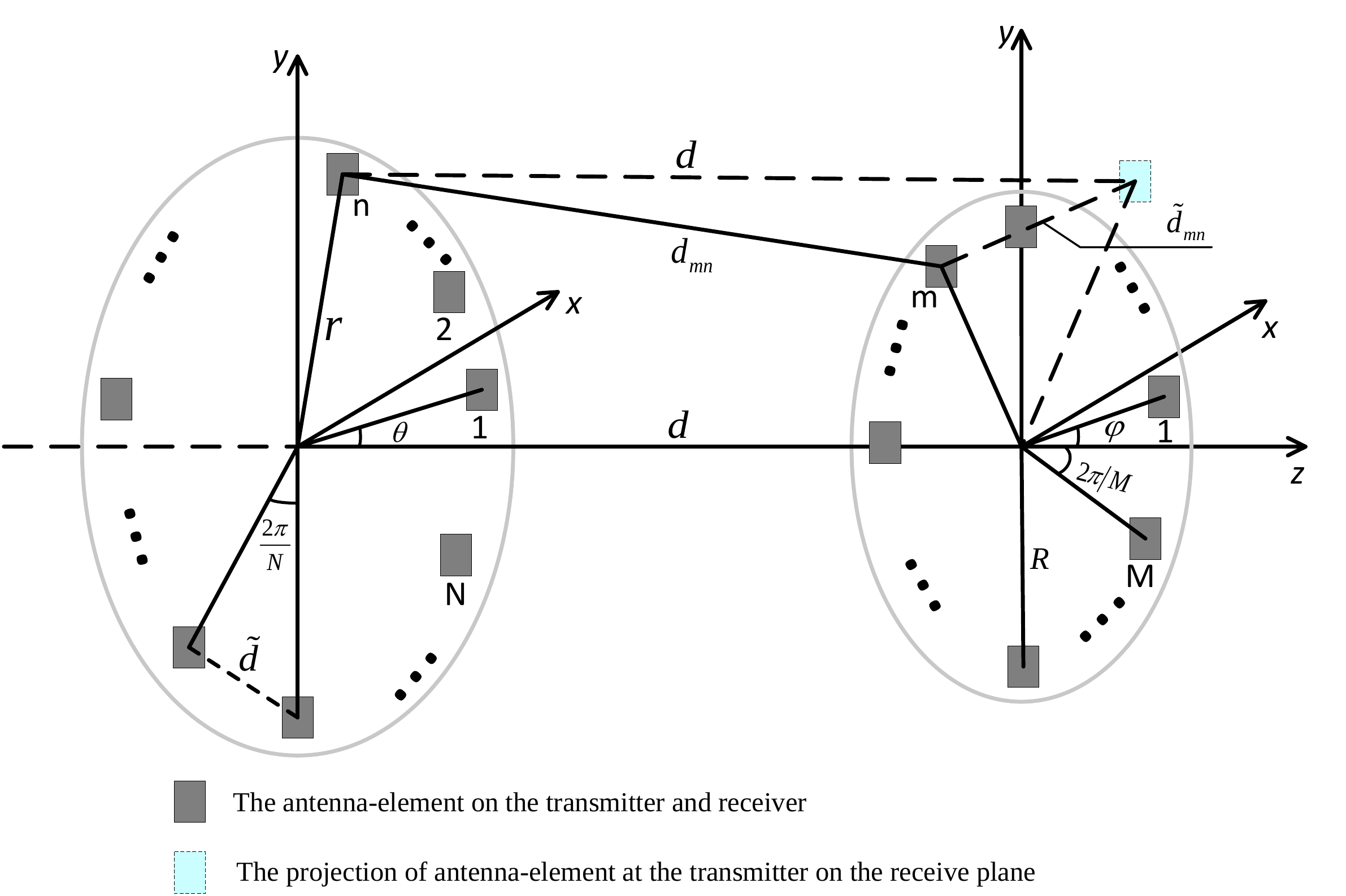}
\caption{The system model for LOS MIMO based wireless communications.}
\label{fig:model}
\end{figure}

Figure~\ref{fig:model} shows the system model for LOS MIMO based wireless communications, where the antennas of transmit and receive UCAs are uniformly around the perimeter of the circle. In this model, the transmit and receive UCAs are coaxial and parallel to each other, i.e., they align well. We denote by $r$ and $R$ the radii of the transmit and receive UCAs, respectively, where $r$ and $R$ can be different. The notation $d$ represents the distance between the center of transmit UCA and the center of receive UCA while $d_{mn}$ represents the distance from the $n$th transmit antenna to the $m$th receive antenna. The notation $\widetilde{d}$ represents the distance between two neighboring antennas. We denote by $\widetilde{d}_{mn}$ the distance between the projection of the $n$th transmit antenna on the receive plane and the $m$th receive antenna. We denote by $\theta$ and $\varphi$ the angles between the phase angle of the first antenna and zero radian corresponding to the transmit and receive UCAs, respectively. There are $N$ and $M$ antennas at the transmit and receive UCAs, respectively.
For the channel independent beamforming scheme, the antennas are fed with the same input signal but with different phase factor. Our proposed channel independent beamforming scheme will be presented in the next section. Note that the antennas are fed with different signals for the traditional LOS MIMO based wireless communications.

\section{Achieving Fast Detection for UCA Based LOS MIMO Communications}\label{sec:signal}
\vspace{-0.0cm}
\subsection{Channel Independent Beamforming Scheme}

We propose to use a channel independent beamforming scheme, where each information symbol is transmitted using all antennas with different phase factors on each antenna. The power of each information symbol is averagely allocated to $N$ transmit antennas.
Let $\left\{s_{1}, s_{2}, \cdots, s_{N}\right\}$ be $N$ information symbols to be sent and $x_{n}$ be the symbol to be transmitted at the $n$th transmit antenna. Then, we propose to use the following modulation:
\begin{eqnarray}\label{eq:transmit_signal}
x_{n} = \sum_{l=1}^{N} \frac{1}{\sqrt{N}} s_{l} e^{j\frac{2\pi (n-1)}{N}(l-1)},\ \ \  n=1, 2, \cdots, N.
\label{OAM_signal}
\end{eqnarray}

According to Eq.~\eqref{OAM_signal}, we can write the vector, denoted by ${\bm x}$, corresponding to the transmit signals at the transmit UCA as follows:
\begin{eqnarray}
{\bm x}={\bm W}{\bm s},
\end{eqnarray}
where ${\bm s}=\left[s_{1}, s_{2}, \cdots, s_{N}\right]^{T}$ is the information symbol vector and the matrix ${\bm W}$ is the standard $N \times N$ IDFT matrix with elements $\left\{\frac{1}{\sqrt{N}}\exp[2\pi (n-1)(l-1)/N]\right\}$, $1 \leq l,n \leq N$.
The expression of the transmit signal is similar to that of the traditional orthogonal frequency-division multiplexing (OFDM)~\cite{ODFM_2009}. But it is different in the sense that the transmission here is along the antenna direction, i.e.,  the $n$th subcarrier signal $x_{n}$ is transmitted in the $n$th transmit antenna instead of the $n$th time slot. In addition, all the $N$ subcarrier symbols are transmitted in one time slot, which means that the symbol rate is $N$, i.e., $N$ symbols per channel use. Also, different from OFDM, the above proposed scheme in Eq.~\eqref{eq:transmit_signal} does not add any CP in the transmission.

\subsection{Channel Model for UCA Based LOS MIMO Antenna Structure}\label{sec:channel}

We denote by $h_{mn}$ the channel gain from the $n$th antenna on the transmit UCA to the $m$th antenna on the receive UCA. Then, $h_{mn}$ can be written as follows~\cite{Goldsmith_book}:
\begin{eqnarray}\label{channel_1}
h_{m,n}=\frac{\beta \lambda e^{-j\frac{2\pi}{\lambda}d_{mn}}}{4\pi d_{mn}},
\end{eqnarray}
where $\beta$ denotes the combination of all the relevant constants, such as, attenuation and phase rotation caused by antennas and their patterns on both sides.
Then, the angle, denoted by $\alpha_{mn}$, between the projection of the $n$th transmit antenna on the receive plane and the $m$th receive antenna is derived as follows:
\begin{eqnarray}
\alpha_{mn}&=&\phi_{n}+\theta-(\psi_{m}+\varphi)\nonumber \\
&=&\frac{2\pi (n-1)}{N}-\frac{2\pi (m-1)}{M}+\theta-\varphi,
\end{eqnarray}
where $\phi_{n}+\theta=2\pi (n-1)/N +\theta$ is the phase angle of the $n$th transmit antenna and $\psi_{m}+\varphi=2\pi(m-1)/M +\varphi$ is the phase angle corresponding to the $m$th receive antenna. Also, $\phi_{n}+\theta$ is the phase angle of the projection corresponding to the $n$th transmit antenna on the receive plane. Then, $\widetilde{d}_{mn}$, which is the distance between the projection of the $n$th transmit antenna on the receive plane and the $m$th receive antenna, is given as follows:
\begin{eqnarray}
\widetilde{d}_{mn}=\sqrt{r^{2}+R^{2}-2rR\cos\left( \alpha_{mn} \right)}.
\end{eqnarray}
Because the direction from the $n$th transmit antenna to the projection of the $n$th transmit antenna on the receive plane is perpendicular to the receive plane, we can derive $d_{mn}$ as follows:
\begin{eqnarray}\label{eq:d}
\hspace{-1cm}d_{mn}\!\!\!&=&\!\!\! \sqrt{d^2+{\widetilde{d}_{mn}}^2}\nonumber \\
\!\!\!&=&\!\!\! \sqrt{d^{2}+r^{2}+R^{2}-2rR\cos\left(\alpha_{mn}\right)}.
\end{eqnarray}
Substituting Eq.~\eqref{eq:d} into Eq.~\eqref{channel_1}, we have
\begin{eqnarray}
\!\!\!\!\!\!h_{m,n}=&&\!\!\!\!\!\!\!\!\! \frac{\beta \lambda}{4\pi \sqrt{d^{2}+r^{2}+R^{2}-2rR\cos\left(\alpha_{mn}\right)} } \nonumber \\
&&\!\!\!\!\!\!\!\!\!\!\!\!\!\! \times \exp\left\{-j\frac{2\pi}{\lambda} \sqrt{d^{2}+r^{2}+R^{2}-2rR\cos\left(\alpha_{mn}\right)}  \right\}\!\!.
\end{eqnarray}
Due to
\begin{eqnarray}
\cos\left(\alpha_{mn}\right)&=&\cos\left[ \frac{2\pi(n-m)}{N}+\theta-\varphi\right] \nonumber \\
&=&\cos\left[ \frac{2\pi (n-m\pm N)}{N}+\theta-\varphi \right],
\end{eqnarray}
if the number of antennas at the transmitter is equal to the number of antennas at the receiver, i.e., $M=N$, matrix ${\bm A}$ with elements $\left[ \cos(\alpha_{mn})\right]$, $ 1 \leq m,n \leq N$, is a circulant matrix when the transmit and receive UCAs are aligned with each other.
Since $h_{m,n}$ depends on the difference between $ \phi_{n}$ and $\psi_{m}$ (i.e., $\cos\left(\phi_{n}-\psi_{m}\right)$) for fixed $d$, $r$, $R$, $\theta$, and $\varphi$, the channel matrix, denoted by ${\bm H}$, can be expressed by
%
\begin{eqnarray}
{\bm H}=\left[\begin{matrix}
h_{1,1}& h_{1,2}& \cdots &h_{1,N} \\[0.3cm]
h_{1,N}& h_{1,1}& \cdots &h_{1,N-1} \\
\vdots &\vdots & \ddots &\vdots \\
h_{1,2}& h_{1,3}& \cdots & h_{1,1} \\
\end{matrix}\right]
\end{eqnarray}
and therefore, is also a circulant matrix.

Because ${\bm H}$ is a circulant matrix, the DFT matrix can diagonalize it:
\begin{eqnarray}
{\bm W}^{*}{\bm H} {\bm W}={\bm \Lambda},
\end{eqnarray}
where
\begin{equation}\label{eq:h_DFT}
{\bm \Lambda}={\rm diag}\left( H_{1,1}, H_{1,2}, \cdots, H_{1,N} \right).
\end{equation}
In Eq.~\eqref{eq:h_DFT}, the sequence $\left\{H_{1,1}, H_{1,2}, \cdots, H_{1,N}\right\}$ is the $N$-point DFT of the sequence $\left\{h_{1,1}, h_{1,2}, \cdots, h_{1,N}\right\}$ corresponding to the first row of the matrix ${\bm H}$, i.e.,
\begin{eqnarray}
H_{1,k}=\sum\limits_{n=1}^{N} h_{1,n} e^{-j\frac{2\pi(n-1)(k-1)}{N}}.
\end{eqnarray}

\vspace{-0.3cm}
\begin{figure}[h]
\subfigure[$\left|H_{1,k}\right|$ for $N=4$.]{\hspace{-0.0cm}
        \includegraphics[width=0.46\linewidth]{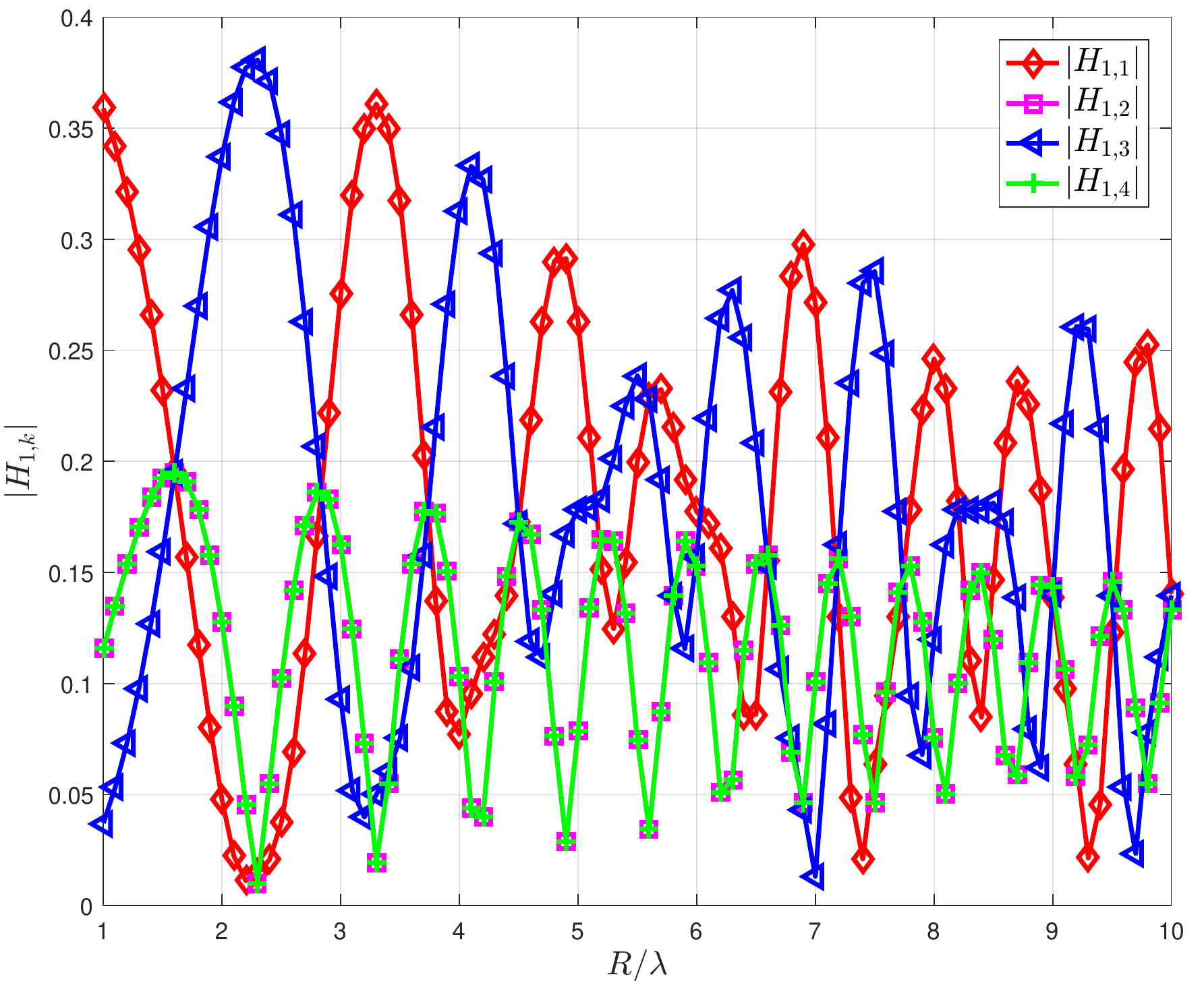}     
        \label{fig:H_k4}
    }
    \subfigure[$\left|H_{1,k}\right|$ for $N=6$.]{\hspace{0.0cm}
        \includegraphics[width=0.46\linewidth]{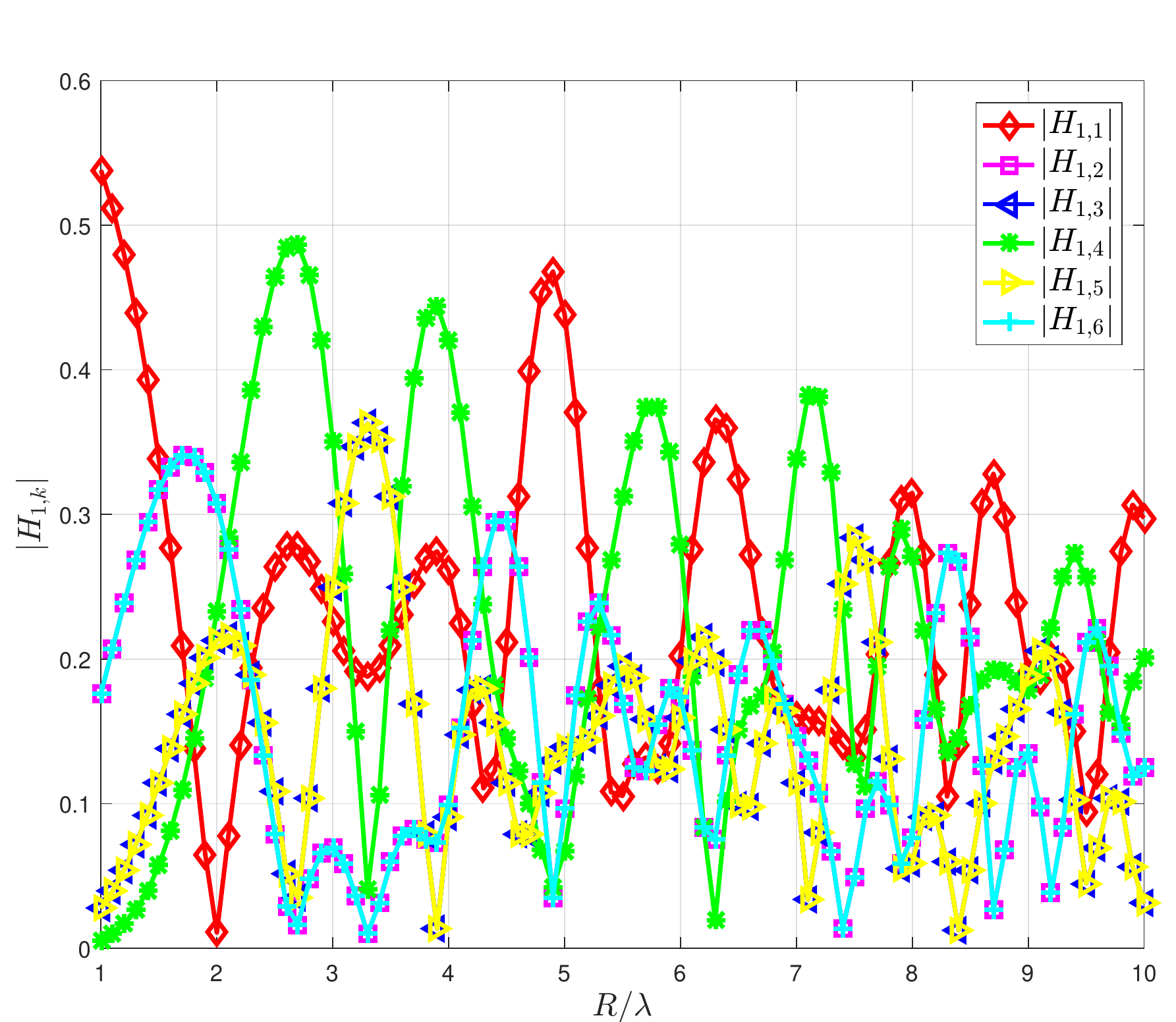}
        \label{fig:H_k6}
    }
        \subfigure[$\left|H_{1,k}\right|$ for $N=8$.]{\hspace{0.0cm}
        \includegraphics[width=0.46\linewidth]{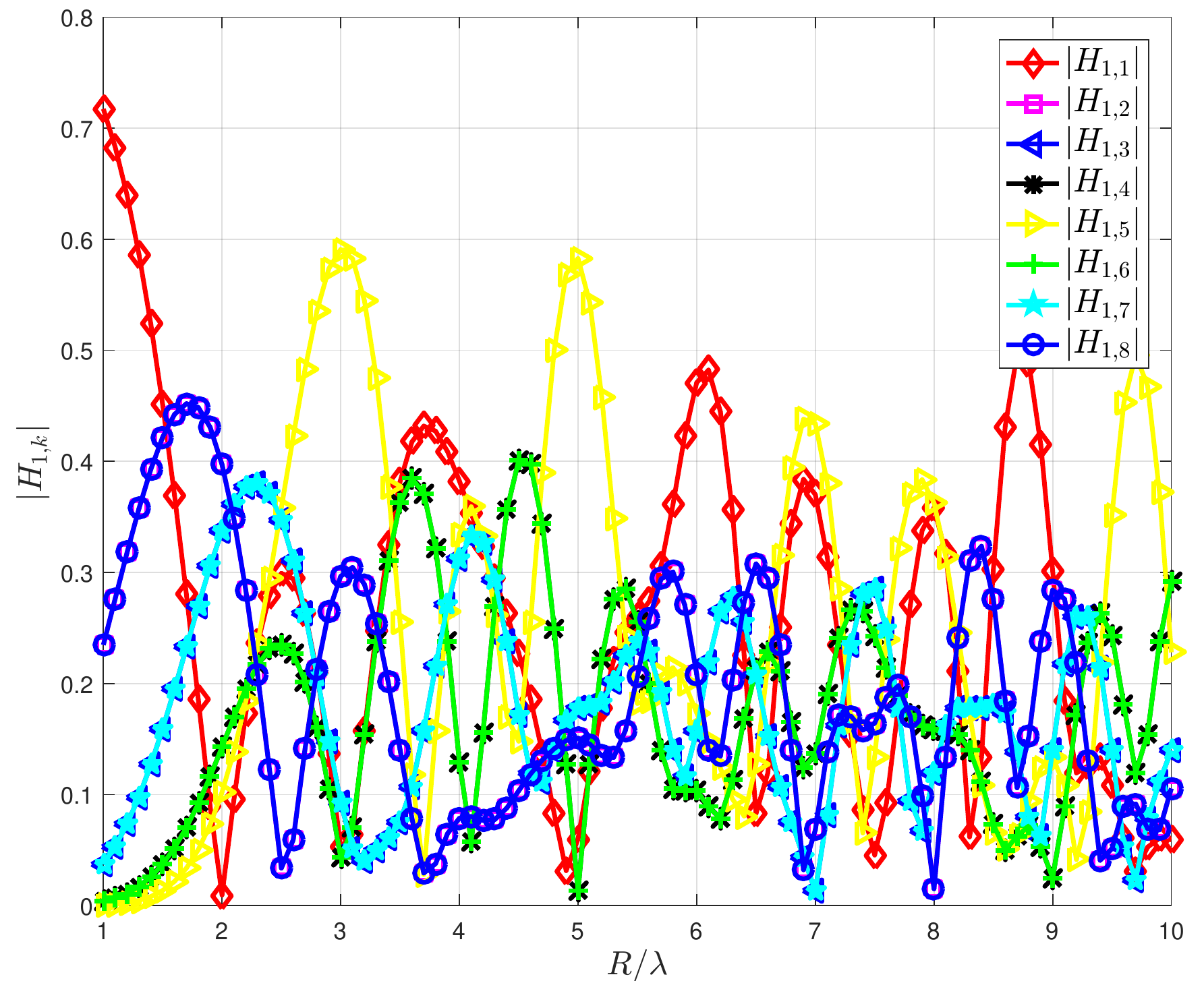}
        \label{fig:H_k8}
    }
        \subfigure[The variances of $\{\left|H_{1,1}\right|,$ $\left|H_{1,2}\right|, \cdots, \left|H_{1,N}\right|\}$ for different numbers of antennas.]{\hspace{0.0cm}
        \includegraphics[width=0.47\linewidth]{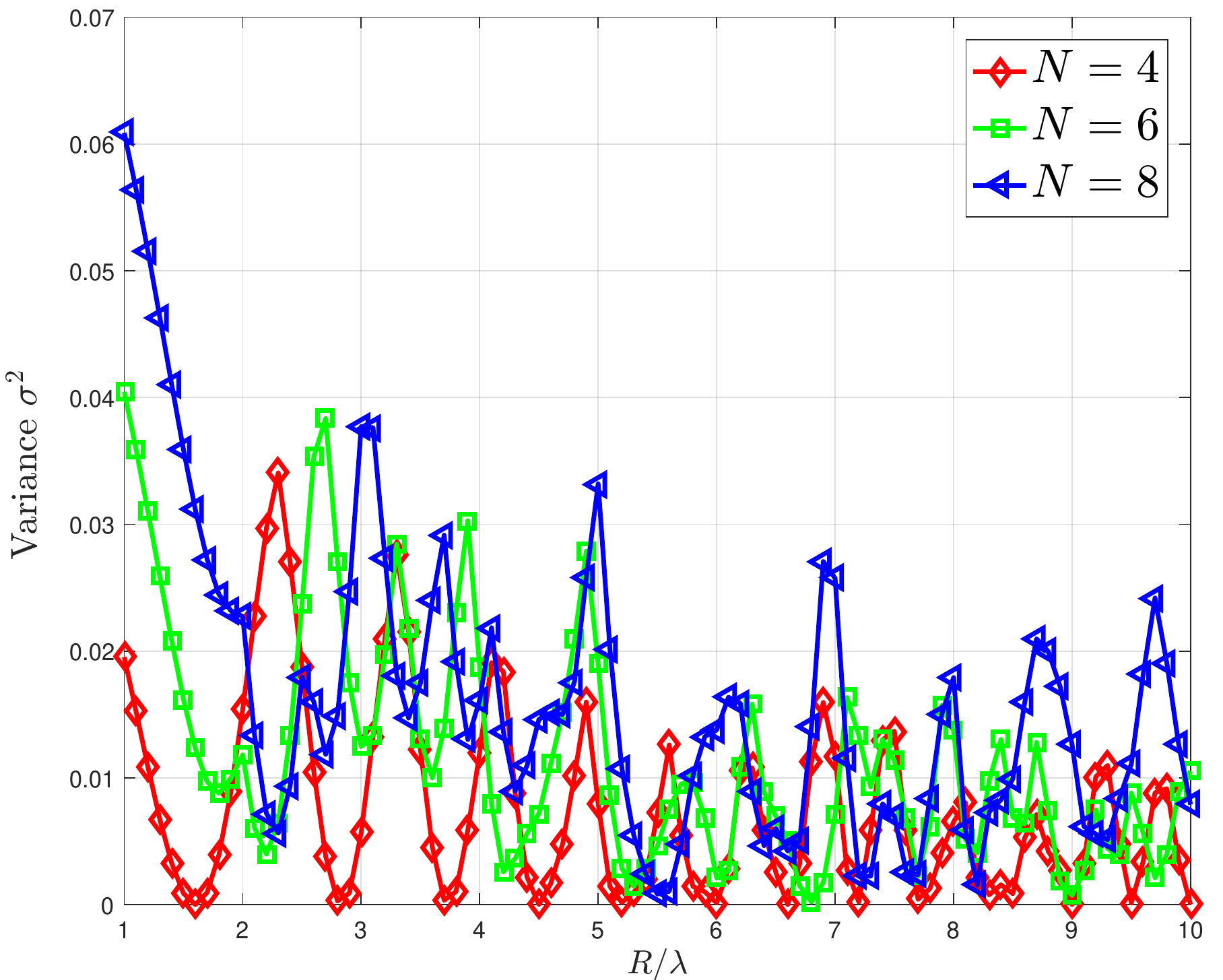}
        \label{fig:varance}
    }
\caption{$\left|H_{1,k}\right|$ and the variances of $\{\left|H_{1,1}\right|, \left|H_{1,2}\right|, \cdots,$ $\left|H_{1,N}\right|\}$ for different numbers of antennas.}\label{fig:mode}
\end{figure}

The variance, denoted by $\sigma^2$, of $\{\left|H_{1,1}\right|, \left|H_{1,2}\right|, \cdots,$ $\left|H_{1,N}\right|\}$ is given as follows:
\begin{eqnarray}
\sigma^2=\frac{1}{N}\sum_{k=1}^{N}\left(\left|H_{1,k}\right|-\frac{1}{N} \sum_{i=1}^{N} \left|H_{1,i}\right|\right)^{2}.
\end{eqnarray}
Similar to the conventional OFDM system, when the variance of $\left|H_{1,k}\right|$ along $k$ is smaller, i.e., $\sigma^2$ is smaller, the channel is better.

\begin{figure}[t]
\centering
\includegraphics[width=0.38\textwidth]{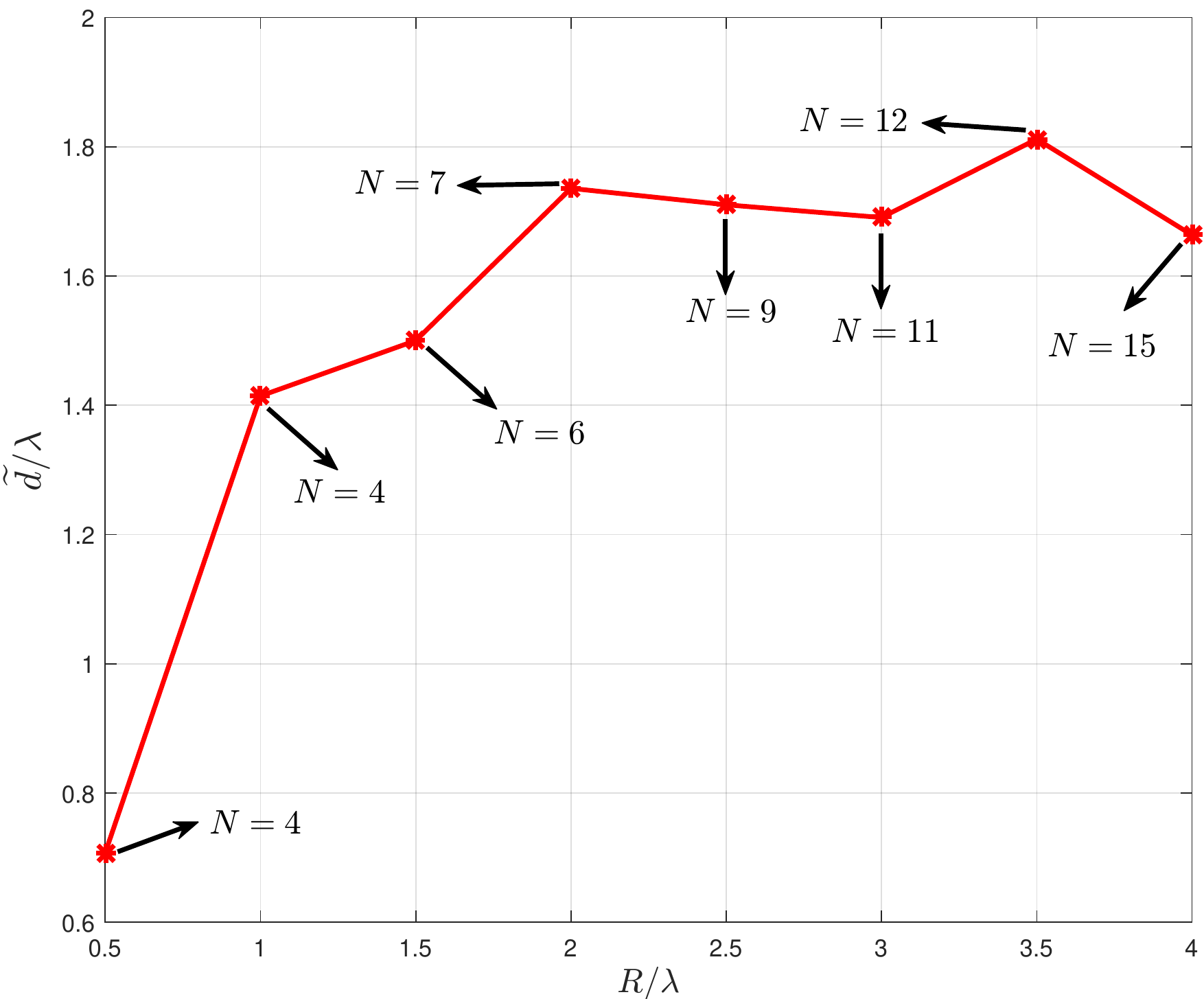}
\vspace{-0.2cm}
\caption{The distance between two neighboring antennas with $\sigma^2<0.01$.}
\vspace{-0.2cm}
\label{fig:distance}
\end{figure}

Figs.~\ref{fig:H_k4},~\ref{fig:H_k6}, and~\ref{fig:H_k8} show $\left|H_{1,k}\right|$, ($k=1, \cdots, N$) for different numbers of antennas. Fig.~\ref{fig:varance} gives the corresponding variances of $\{\left|H_{1,1}\right|, \left|H_{1,2}\right|,\cdots, \left|H_{1,N}\right|\}$. 
We set $\beta=4\pi$, $\theta=\varphi=0$, $R=r$, and $d=10\lambda$. The variances for different numbers of antennas decrease as the radius increases. This is because the distance between two neighboring antennas increases as the radius increases.


Fig.~\ref{fig:distance} gives the normalized distance by wavelength, i.e., $\widetilde{d}/\lambda$, between two neighboring antennas with $\sigma^2<0.01$. As $R$ increases, the maximum possible number of antennas increases given $\sigma^2<0.01$. $\widetilde{d}$ is roughly $1.7\lambda$ when $R$ is larger than $2\lambda$, corresponding to the optimal $N$ with $\sigma^2<0.01$. Thus, we can follow this setup to avoid that the channels are too correlated.

\vspace{-0.0cm}
\subsection{Fast Symbol-Wise ML Detection Scheme}\label{sec:receive}

The receive signal vector for the LOS MIMO system, denoted by ${\bm y}$, can be derived as follows:
\begin{eqnarray}\label{eq:receive_OAM}
{\bm y}={\bm H} {\bm x} + {\bm z}={\bm H} {\bm W}{\bm s}+ {\bm z},
\end{eqnarray}
where ${\bm y}=[y_{1}, \cdots, y_{m}, \cdots, y_{N}]^{T}$ and ${\bm z}=[z_{1}, \cdots, z_{m}, \cdots, z_{N}]^{T}$ with $y_{m}$ and $z_{m}$ representing the receive signal and the noise of the $m$th receive antenna.



%

To obtain the transmitted information symbols, we multiply the matrix ${\bm W }^{*}$ to the received signal vector ${\bm y}$:
\begin{eqnarray}
{\bm W }^{*}{\bm y}&=& {\bm W }^{*} {\bm W \bm \Lambda}{\bm s}+ {\bm W }^{*}{\bm z}\nonumber \\
&=&{\bm \Lambda}{\bm s}+ {\bm W }^{*}{\bm z},
\end{eqnarray}
which is the same as taking the $N$-point DFT along the receive antenna direction. Because the DFT matrix ${\bm W }^{*}$ is a unitary matrix, the noise statistics dose not change. Denote $\widetilde{{\bm y}}={\bm W }^{*}{\bm y}=[\widetilde{y}_{1}, \widetilde{y}_{2}, \cdots, \widetilde{y}_{N}]^{T}$. Then, we can obtain an estimate, denoted by $\widehat{{\bm s}}$, of the transmitted information symbol vector ${\bm s}$ via the ML detection as follows:
\begin{eqnarray}\label{eq:rece_detection}
\!\!\!\!\!\!\!\!\!\!\!\!\!\!\!\widehat{{\bm s}}\!\!\!\!&=&\!\!\!\!\arg \min_{{\bm s}\in {\bm \Omega}^{N}} \left\| \widetilde{{\bm Y}} -  {\bm \Lambda}{\bm s} \right\|^2\nonumber \\
\!\!\!\!&=&\!\!\!\! \arg \min_{{\bm s}\in {\bm \Omega}^{N}} \sum_{l=1}^{N}\left| \widetilde{y}_{l} -   H_{1,l} s_{l} \right|^2\!\!\nonumber \\
\!\!\!\!\!\!&=&\!\!\!\!\!\! \left[\arg \min_{\!\!\! s_{1}\in {\bm \Omega}}\left| \widetilde{y}_{1} -  H_{1,1} s_{1} \right|^2\!\!, \cdots, \arg \min_{ \!\!\!s_{N}\in {\bm \Omega}}\left| \widetilde{y}_{N} -  H_{1,N} s_{N} \right|^2\right]^{T}\!\!\!\!\!, \nonumber \\
\end{eqnarray}
where ${\bm \Omega}$ is a signal constellation of size $K$.
The last equality in Eq.~\eqref{eq:rece_detection} is for an uncoded system where the information symbols $[s_1, ..., s_N]$ are independent.

For the traditional LOS MIMO system with $N$ transmit antennas, each antenna transmits one symbol. Thus, the corresponding traditional ML detection scheme is given as follows:
\begin{eqnarray}
\widehat{{\bm s}}=\arg \min_{{\bm s}\in {\bm \Omega}^{N}} \left\|{\bm y}- {\bm H}{\bm s} \right\|^{2}.
\end{eqnarray}


Note that our proposed channel independent beamforming is a basic modulation for UCA based LOS MIMO. Similar to the conventional OFDM for ISI channels, if the channel state information is known at the transmitter, further precoding can be employed across the subcarriers to improve the performance. This is particularly important when $N$ is large and the channels become too correlated.

\subsection{BER and Computational Complexity Analyses}

The BER, denoted by $P_{e}$, corresponding to binary phase shift keying (BPSK) for the LOS MIMO system using the channel independent beamforming can be easily derived from Eq.~\eqref{eq:rece_detection} as follows~\cite{LOS_channel_2007}:
\begin{equation}
P_{e}= \frac{1}{N}\sum\limits_{l=1}^{N} \frac{1}{2} {\rm erfc}\left( \frac{\left|H_{1,l}\right|^2 \left|s_{l}\right|^2}{N \omega^2 } \right),
\end{equation}
where ${\rm erfc} (x)= \frac{2}{\sqrt{\pi}}\int_{x}^{\infty} e^{-t^2}dt$ and $\omega^2$ represents the variance of received noise.

\renewcommand\arraystretch{2}
\begin{table}[h]
\begin{center}
\caption{Complexity Comparison}
\label{tab:para}
\begin{tabular}{ m{1.6cm}<{\centering}|  m{2.5cm}<{\centering}|  m{2.5cm}<{\centering}}
  \hline
  \rowcolor[gray]{.9}
  Detection scheme & The number of complex additions & The number of complex multiplications \\
  \hline
  \hline
  \rowcolor[gray]{.9}
  Fast symbol-wise ML  & $N\log_{2}(N)+NK$ & $\frac{N}{2}\log_{2}(N)+NK$ \\\hline
  \rowcolor[gray]{.9}
  Traditional ML  & $N^2K^N$ & $(N^2+N)K^N$ \\\hline
\end{tabular}
\end{center}
\vspace{-10pt}
\end{table}

The computational complexities of our proposed scheme with the fast symbol-wise ML detection and the traditional scheme with the ML detection are listed in Table~\ref{tab:para}. In Table~\ref{tab:para}, $K$ represents the size of modulation alphabet. We can observe that the numbers of complex additions and complex multiplications corresponding to the fast symbol-wise ML detection are much smaller than those corresponding to the traditional ML detection, respectively. 

%
%
%

\section{Numerical Results}
\begin{figure}[b]
\centering
\includegraphics[width=0.37\textwidth]{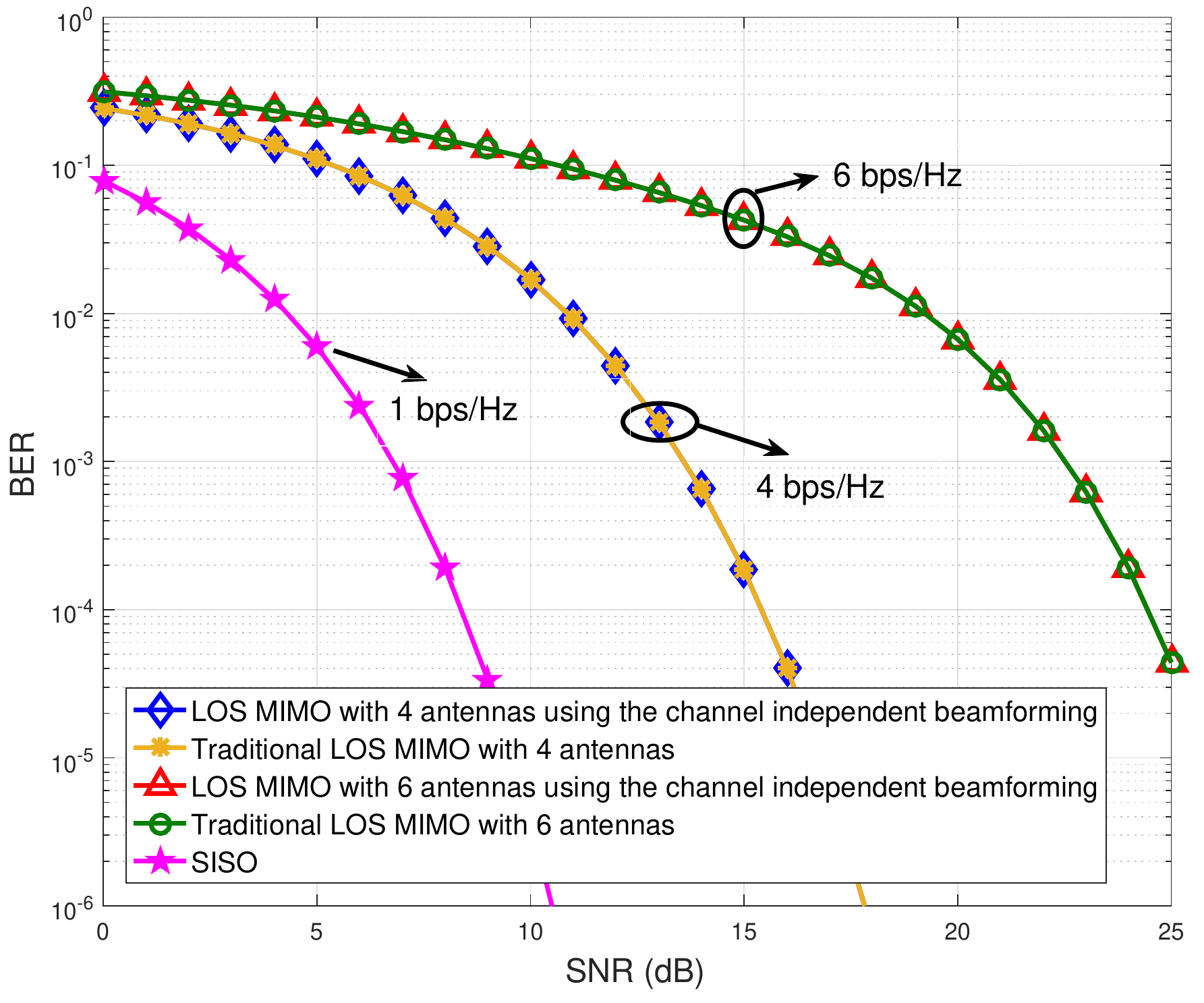}
\vspace{-0.1cm}
\caption{BERs for the LOS MIMO system using the channel independent beamforming and the traditional LOS MIMO system (BPSK).}
\label{fig:Bit_error}
\end{figure}
We numerically evaluate the LOS MIMO system using the channel independent beamforming. We compare the BER for the LOS MIMO system using the channel independent beamforming and the BER of the traditional LOS MIMO system, where we set $\beta=4\pi$, $\theta=\varphi=0$, $\lambda=0.1$ m, $R=r=0.4$ m, and $d=1.2$ m. 

Figure~\ref{fig:Bit_error} shows the BERs for the LOS MIMO system using the channel independent beamforming and the traditional LOS MIMO system. We can observe that the BERs for the two schemes are the same. This is because our proposed channel independent beamformers at both transmit and receive sides are IDFT and DFT, respectively, that are unitary transforms and therefore, they do not change the channel properties. The BER increases as the number of antennas increases, while the transmission throughput increases as well. 
Note that, although their BER performances are the same, as it is shown in Table~\ref{tab:para}, the computational complexity of our proposed scheme is much lower.

\section{Conclusions} \label{sec:conc}
In this letter, we proposed a channel independent beamforming and the corresponding fast symbol-wise ML detection, where the transmit and the receive UCAs are aligned with each other. It has symbol rate $N$, i.e., $N$ symbols per channel use. To our best knowledge, it is the first and the only MIMO system so far with both symbol rate $N$ and the fast symbol-wise ML detection for $N$ transmit antennas. Our proposed channel independent beamformers are simply IDFT and DFT, respectively, at the transmitter and the receiver. We compared BER and the number of computations corresponding to the LOS MIMO system using the channel independent beamforming and the corresponding fast symbol-wise ML detection with those of the traditional LOS MIMO system. While the BER of the LOS MIMO system using the channel independent beamforming is the same as that of the traditional LOS MIMO system, the numbers of complex additions and complex multiplications are much smaller than those of the traditional LOS MIMO detection scheme.


\end{document}